\documentclass[11pt,twoside]{article}
\usepackage{asp2010}

\newcommand{\Ms}{{\cal M}_{\rm s}}
\newcommand{\Msun}{M_\odot}
\newcommand{\NH}{N_{\rm H}}
\newcommand{\pcc}{{\rm ~cm}^{-3}}
\newcommand{\psc}{{\rm ~cm}^{-2}}

\resetcounters

\begin{document}

\title{Molecular Cloud Evolution}
\author{Enrique V\'azquez-Semadeni$^1$
\affil{$^1$Centro de Radioastronom\'\i a y Astrof\'\i sica, UNAM, Campus
Morelia, P.O. Box 3-72, Morelia, Michoac\'an, 58089, M\'exico}}

\begin{abstract}
I describe the scenario of molecular cloud (MC) evolution that has
emerged over the past decade or so. MCs can start out as cold atomic
clouds formed by compressive motions in the warm neutral medium (WNM) of
galaxies. Such motions can be driven by large-scale instabilities, or by
local turbulence. The compressions induce a phase transition to the cold
neutral medium (CNM) to form growing cold atomic clouds, which in their
early stages may constitute thin CNM sheets. Several dynamical
instabilities soon destabilize a cloud, rendering it turbulent.
For solar neighborhood conditions, a cloud is coincidentally expected
to become molecular, magnetically supercritical, and gravitationally
dominated at roughly the same column density, $N \sim 1.5 \times 10^21
\psc \approx 10 \Msun$ pc$^{-2}$. At this point, the cloud begins to
contract gravitationally. However, before its global collapse is
completed ($\sim 10^7$ yr later), the nonlinear density fluctuations
within the cloud, which have shorter local free-fall times, collapse
first and begin forming stars, a few Myr after the global contraction
started. Large-scale fluctuations of lower mean densities collapse
later, so the formation of massive star-forming regions is expected to
occur late in the evolution of a large cloud complex, while scattered
low-mass regions are expected to form earlier. Eventually, the local
star formation episodes are terminated by stellar feedback, which
disperses the local dense gas, although more work is necessary to
clarify the details and characteristic scales of this process.
\end{abstract}

\section{Introduction} \label{sec:intro}

The molecular gas in the Galaxy exists in giant complexes ({\it giant
molecular clouds}, or GMCs) of masses $\sim 10^{5-6} \Msun$, sizes of
several tens of parsecs, and mean densities $n \sim 100 \pcc$
\citep[see, e.g., the review by][]{1993prpl.conf..125B}, and contain a
large amount of substructure (parsec-scale {\it
clumps} of densities $n \sim 10^3 \pcc$, and sub-parsec-scale {\it
cores} of densities $n > 10^4 \pcc$). Molecular
clouds (MCs) contain roughly half the gaseous mass in the 
Galaxy, and are the sites of all present-day star formation (SF) in the
Galaxy. Thus, the study of their origin and evolution is crucial for our
understanding of SF, besides the importance, on its own right, of
understanding this fundamental component of the ISM.

The seminal paper by \citet{1954BAN....12..177O} postulated the
existence of a cycle (now known as the {\it Oort cycle}), in which the
expanding HII regions around newly formed massive stars form shells of
cold gas around them.  The shells subsequently fragment and produce a
population of cloudlets, which then grow by coagulation until they
become gravitationally unstable, at which point they proceed to collapse
and form a new generation of stars, starting the cycle all over
again. In this model, which was later developed by
\citet{1965ApJ...142..568F}, clouds were assumed to grow exclusively by
coagulation of randomly-moving small cloudlets. The same cloud growth
process was assumed in later models, such as that by
\citet{1980ApJ...238..158N}, which differed from the Oort model mainly
in that the driving agent was considered to be winds from low-mass,
T-Tauri stars rather than the ionizing radiation from massive stars, and
the model by \citet{1989ApJ...345..782M} for the SF rate (SFR) regulated
by the background photo-ionizing radiation. The coagulation process
implied very long ($\gtrsim 10^8$ yr) cloud growth times
\citep{1979ApJ...229..578S, 1979ApJ...229..567K}, which were however
ruled out on the basis of observational evidence by
\citet{1980ApJ...238..148B}. These authors proposed instead that MCs
form and grow by a \citet{1966ApJ...145..811P} instability triggered
by spiral-arm shocks, and have lifetimes $\sim 10^7$ yr.

Ever since the times of those early models, MCs have been an odd
component of the ISM. Because they are known to be strongly
self-gravitating \citep{1981MNRAS.194..809L, 1988ApJ...326L..27M} and at
significantly higher thermal pressures than the mean ISM pressure
\citep[e.g.,][]{1978ApJ...225..380M}, they did not fit in
thermal-pressure balance models of the ISM, such as that by
\citet{1977ApJ...218..148M}. Instead, they have traditionally been
considered to be in approximate virial equilibrium
\citep{1981MNRAS.194..809L, 1988ApJ...326L..27M}, supported against
their self-gravity by either the magnetic field \citep[the so-called
``standard'' model of magnetically regulated SF; see, e.g., the reviews
by][]{1987ARA&A..25...23S, 1991psfe.conf..449M}, or by turbulence driven
by stellar feedback \citep[the so-called ``turbulent'' model of SF; see
the reviews by][]{2000prpl.conf....3V, 2004RvMP...76..125M,
2007ARA&A..45..565M, 2007prpl.conf...63B}. In both cases, the
gravitational contraction was assumed to be halted by either of the two
mechanisms, and the clouds were assumed to reach near virial
equilibrium. In the last decade or so, however, the paradigm about the
formation, evolution and structure of MCs has changed significantly, and
in the remainder of this review I will discuss this emerging new view.


\section{Birth and infancy} \label{sec:formation}

\subsection{Observational and numerical evidence on the clouds' origin}
\label{sec:evidence} 

As mentioned above, GMCs and their substructures had traditionally been
thought to be in virial equilibrium.  However, recent observations of
GMCs in the LMC \citep{2009ApJS..184....1K} suggest that the clouds are
undergoing an evolutionary process, in which both their mass and their
SF activity increase in time, going in $\sim 25$ Myr from masses $M \sim
10^{4.5-5} \Msun$ and virtually no massive-SF to $M \sim 10^{5.5-6}
\Msun$ and a population of clusters and HII regions. This is consistent
with the conclusion by \citet{2003ApJS..149..343E} that the GMCs in M33
are being assembled rapidly from the atomic component, with a prompt
onset of SF.

Similar conclusions had been reached previously from numerical
studies. Numerical simulations of the ISM at the kpc scale with
turbulence driven by stellar feedback \citep[supernova explosions,
expanding HII regions;][]{1980ApJ...239..173B, 1988ApJ...328..427C,
1993ApJ...413..137R, 1995ApJ...440..634R, 1995ApJ...441..702V,
1995ApJ...455..536P, 1999ApJ...514L..99K, 1999ApJ...527..285B,
2000MNRAS.315..479D, 2004A&A...425..899D, 2005A&A...436..585D} showed
that compressive motions driven by large-scale gravitaional
instabilities in the diffuse ISM or by the global turbulence
are able to form clouds on short timescales, essentially given by the
turbulent crossing time accross the distance necessary to collect the
material that eventually reaches the cloud. This is facilitated by the
presence of cooling, which renders the medium highly compressible, even
when no thermal instability (cf.\ \S \ref{sec:instab}) is present
\citep{1996ApJ...473..881V}. Within such a dynamic scenario,
\citet{1999ApJ...515..286B} remarked that the clouds should not be
considered as isolated objects, because a significant mass flux is
expected to exist accross their boundaries, since the clouds are being
assembled from material from the outside.

\subsection{The physical processes}
\label{sec:phys_proc} 

\subsubsection{Instabilities galore!} \label{sec:instab}

The scenario of cloud assembly by convergent motions in the diffuse ISM
(either driven by turbulence or by large-scale instabilities) was
formulated analytically by \citet{2001ApJ...562..852H}, who pointed out
that the column density of cold atomic hydrogen necessary for H$_2$ and
CO molecules to form is \citep{1988ApJ...334..771V, 1998ARA&A..36..317V} 
\begin{equation}
\NH \sim 1\hbox{--}2 \times 10^{21} \psc,
\label{eq:N_molec_form}
\end{equation}
corresponding to $A_V \sim 0.5$--1, is very similar to the value
necessary for the same gas to become gravitationally unstable \citep[see
also][]{1986PASP...98.1076F}, and to the column density necessary for
the gas to become magnetically supercritical, at a typical interstellar
magnetic field strength of $B\sim 5 \mu$G. This implies that when an
initially atomic cloud is assembled by a convergent velocity field in
the diffuse ISM, it should become molecular, self-gravitating,
and magnetically supercritical at roughly the same time. We now discuss
this phenomenology in some more detail.

A fundamental physical ingredient aiding the formation of dense atomic
and molecular clouds is thermal instability \citep{1965ApJ...142..531F},
which is a consequence of the various radiative heating and
cooling processes operating on the atomic ISM
\citep[for a modern discussion, see][]{1995ApJ...443..152W}. The atomic
ISM is subject to the so-called {\it isobaric} mode of this instability:
for densities in the range $\sim 1$--10 $\pcc$ ($T \sim 5000$--500 K),
the gas {\it loses thermal pressure} upon a dynamic compression
\citep[for a pedagogical discussion, see the review
by][]{2009arXiv0902.0820V}, implying that after the compression it will
be underpressured with respect to its surroundings, and will continue to
be squeezed by them until it exits the thermally unstable range, at
which point any further compression again increases its thermal
pressure, until it eventually reaches pressure equilibrium with its
surroundings, but at a much higher density and lower temperature. This
is the basis of the well-known two-phase model of the atomic ISM
\citep{1969ApJ...155L.149F}, originating the warm-diffuse and cold-dense
phases of this ISM component, which are respectively referred to as the {\it
warm} and {\it cold neutral media} (in turn, respectively, WNM and CNM).

The above discussion assumes that the gas is in the thermally unstable
range to begin with. However, \citet{1999A&A...351..309H} showed that
transonic compressions (i.e., of Mach number $\Ms \gtrsim 1$) in the
{\it stable} warm phase can nonlinearly trigger a transition to the cold
phase, so that cold clouds can be formed out of the stable WNM in the
presence of transonic turbulence, which is indeed observed in this
medium \citep{1987ASSL..134...87K, 2003ApJ...586.1067H}. Alternatively,
large-scale ($\gtrsim 1$ kpc) instabilities in the diffuse medium, such
as gravitational \citep[e.g.,][]{1994ApJ...433...39E}, Parker
\citep{1966ApJ...145..811P} or magneto-Jeans
\citep{2002ApJ...581.1080K}, can provide the driving forces for these
motions.

The gas cooled and compressed by this process to form a cloud is subject
to a large number of dynamical instabilities. It has been long been
known that compressed layers formed by the collision of gas streams are
nonlinearly unstable, meaning that the layers become turbulent when the
colliding flows have sufficiently large velocities
\citep{1986ApJ...305..309H, 1992ApJ...386..265S, 1994ApJ...428..186V}.
This process is known as the {\it nonlinear thin-shell instability}
(NTSI). Furthermore, the presence of cooling lowers the required inflow
velocities to destabilize the layers \citep{2005A&A...438...11P}.
Finally, the Kelvin-Helmholz and Rayleigh-Taylor instabilities are also
expected to operate during the formation of a cloud. The interplay of
all these instabilities has been investigated numerically by
\citet{2005ApJ...633L.113H, 2006ApJ...648.1052H}. In summary, {\it
convergent motions in the WNM are expected to produce turbulent CNM
clouds}.

\subsubsection{Evolution of the mass-to-magnetic flux ratio}
\label{sec:M2FR}

Another crucial ingredient in MC dynamics is the magnetic field, and in
particular, the mass-to-magnetic flux ratio (M2FR). As it is well known,
the magnetic field can support a cloud against the latter's self-gravity
if the M2FR, or, equivalently, the ratio of column density to field
strength, exceeds some critical value which, for a cylindrical
configuration, is given by \citep{1978PASJ...30..671N}

\begin{equation}
\Sigma/B \approx (4 \pi^2 G)^{-1/2},
\label{eq:NN78}
\end{equation}
where $\Sigma$ is the mass column density. We denote by $\mu$ the value
of the M2FR normalized to the critical value. Thus, a {\it magnetically
subcritical} cloud has $\mu < 1$, and can be supported by the magnetic
field, while a {\it supercritical} one has $\mu > 1$, and cannot be
magnetically supported \citep[e.g.,][]{1987ARA&A..25...23S,
1991psfe.conf..449M}. In the ``standard'' model of SF, most clouds were
assumed to be strongly magnetically subcritical, and thus globally
supported by the field. SF was thought to occur on long timescales and
involving small fractions of the clouds' mass because, in the dense
cores, the process known as {\it ambipolar diffusion} (AD) allows a
redistribution of the magnetic flux, so that they can continue to
contract quasi-statically, until their M2FR finally becomes
supercritical, and then the cores collapse. 

It is a very common practice to assume that the M2FR is a conserved
quantity as long as the flow behaves ideally (i.e., AD is negligible).
After all, for an isolated cloud of fixed mass, the mass is constant by
construction, and the magnetic flux is conserved by the flux-freezing
condition \citep[see, e.g.][]{1992phas.book.....S}. However, the M2FR
refers to the mass within flux tubes, and in general, field lines do not
end at the ``edge'' of a cloud, but rather continue out to arbitrarily
long distances. 
In fact, it is well possible that magnetic field lines circle around the whole
Galaxy. Now, eq.\ (\ref{eq:NN78}) implies that a flux tube is
supercritical beyond an {\it accumulation length} given by
\citep{1985prpl.conf..320M, 2001ApJ...562..852H}
\begin{equation}
L_{\rm acc} \approx 470 \left(\frac{B}{5 \mu{\rm G}} \right) \left(\frac{n}{1
\pcc} \right)^{-1}~{\rm pc},
\label{eq:acc_length}
\end{equation}
so that, in principle, the entire ISM is supercritical, at least near
the midplane. However, in the case of a forming CNM cloud, what is
relevant is the M2FR {\it of the dense gas that makes up the cloud},
since the cloud is up to $100\times$ denser than its surroundings, and
thus it is the main source of the self-gravity that the field has to
oppose. Thus, in this problem, natural boundaries for up to where to
measure the M2FR are provided by the bounding surface of the dense gas.

Accumulation of material is {\it not} opposed by the magnetic field if it
occurs along field lines, and so this can occur freely in
the ISM. When compressions occur at an angle with the field, it has been
shown by \citet{2000A&A...359.1124H} that, up to a certain angle that
depends on the Mach number of the inflows and the field strength, the
inflow is reoriented along the field lines, and it behaves as if the
compression were parallel to the lines. Beyond that angle, the
compression behaves essentially as if it were perpendicular to the
field, and the flow bounces off, not forming any cloud.  Thus, in what
follows, we consider the case of a cloud forming along field
lines. Equation (\ref{eq:NN78}) can be rewritten as
\begin{equation}
N_{\rm cr} = 1.45 \times 10^{21} \left(\frac{B}{5 \mu {\rm G}}
\right) \psc,
\label{eq:Sigma_crit}
\end{equation}
implying that the column density for a cloud to become supercritical is
very similar to that for molecule formation to begin, and for the cloud
to become gravitationally unstable (cf.\ eq.\ \ref{eq:N_molec_form}).
Thus, we can infer that, {\it as a cloud forms and grows out of a
compression in the WNM, it should become molecular, supercritical, and
gravitationally unstable at roughly the same time.}

Figure \ref{fig:mu_evol} shows the evolution of the M2FR in two
numerical simulations of cloud formation by V\'azquez-Semadeni et al.\
(2011 in prep.; see also \citeauthor{2009MNRAS.398.1082B}
\citeyear{2009MNRAS.398.1082B}), illustrating the growth of
the M2FR. In these simulations, two cylindrical streams of radius 32 pc
and length 112 pc each, are set to collide against each other at the
center of a 256-pc numerical box along the $x$-direction, so that the
cloud is a thin cylindrical layer perpendicular to the inflows. The M2FR
is measured for each line of sight (LOS) perpendicular to the cloud,
which is defined as the cylindrical volume of radius 32 pc and length 20
pc centered at the numerical box center. For each LOS, the M2FR is
measured as the ratio $\Sigma/B_{\|}$, where $B_{\|}$ is the field
component parallel to the LOS and $\Sigma$ is the mass column density
along each LOS. Thus, these measurements constitute {\it upper limits}
to the real M2FR (for a detailed discussion, see V\'azquez-Semadeni et
al.\ 2011). The simulations have mean magnetic field strengths of 3 and
4 $\mu$G, and $\mu = 0.91$ and 0.68, respectively. That is, both
simulations are globally subcritical, implying that no subregion of it
can be supercritical \citep{2005ApJ...618..344V} as long as the flow
remains ideal. Values of $\mu$ greater than those of the whole box
indicate that the measured values overestimate the actual
$\mu$. Nevertheless, for a real cloud not limited by the box size, the
magnetic criticality is expected to eventually become $>1$.

\articlefigure{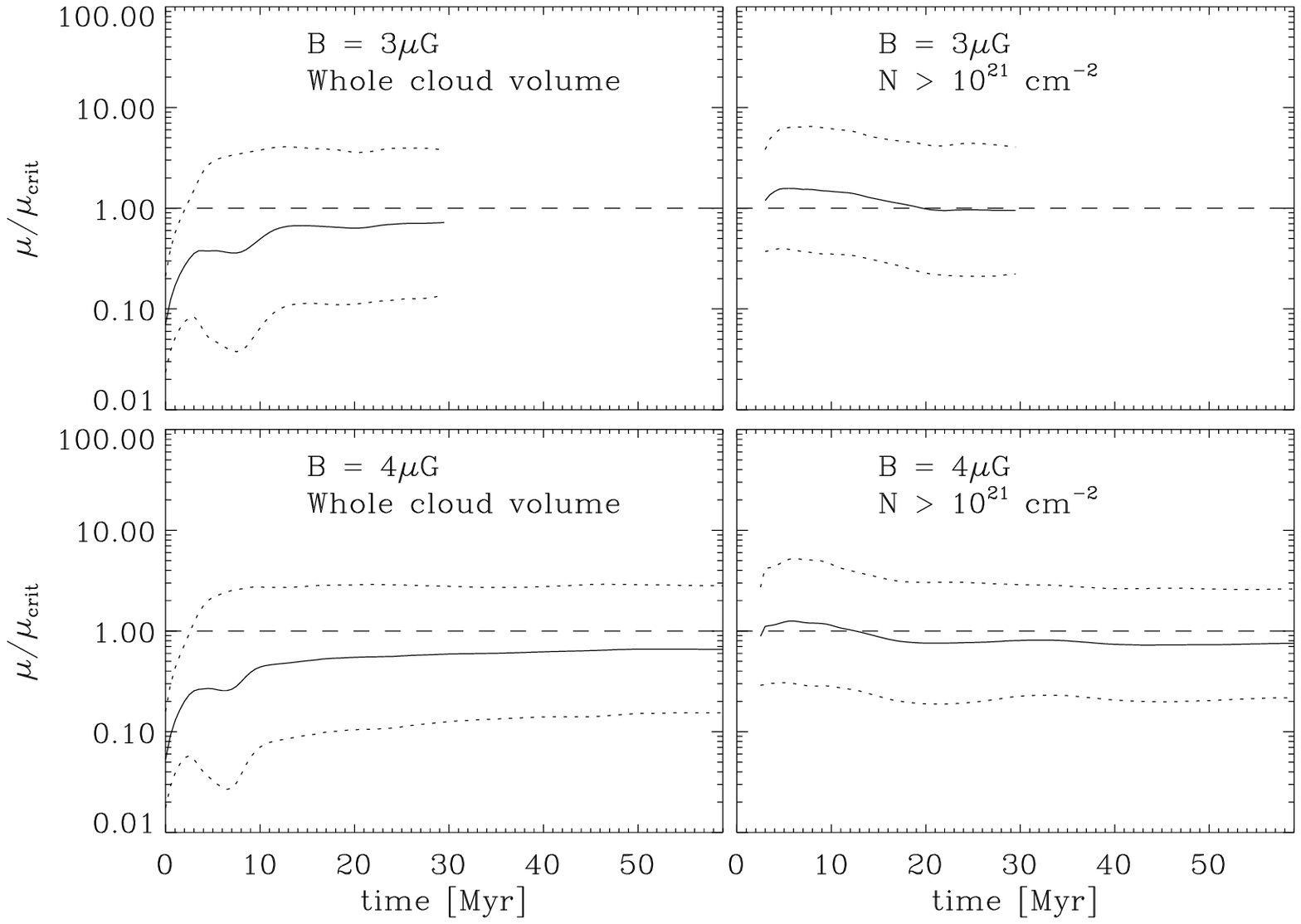}{fig:mu_evol}{Evolution of the
normalized M2FR, $\mu$, for
two numerical simulations by V\'azquez-Semadeni et al.\
(2011, in prep.). See text for details on the simulations. The {\it top
panels} refer to the simulation with $\langle B \rangle = 3 \mu$G, while
the {\it bottom panels} refer to the simulation with $\langle B \rangle = 4
\mu$G. The {\it left panels} refer to measurements performed on the
entire ``cloud'' volume, while the {\it right} panels refer to
measurements perform only on those lines of sight for which the number
column density satisfies $N > 10^{21} \psc$. The {\it solid lines}
denote the average of $\mu$ over all LOSs, while the {\it dotted lines}
denote the $3 \sigma$ deviations from the mean. The growth of $\mu$ from
very small values to near unity during the first 10 Myr of evolution is
clearly seen, especially in the plots for the whole cloud.}

These conclusions are consistent with
observational results showing that the CNM clouds are in general
magnetically subcritical \citep[][sec. 7]{2005ApJ...624..773H}, while
molecular structures appear to be critical or supercritical in general
\citep{2001ApJ...554..916B, 2008ApJ...680..457T}

\subsection{The early stages: thin CNM sheets} \label{sec:early}

The early stages of cloud formation were investigated analytically and
numerically by \citet{2006ApJ...643..245V}, determining the structure
and physical conditions (density, temperature, and expansion velocity of
the phase transition front) in the incipient cloud as a function of the
Mach number of the converging gas streams, before the dynamical
instabilities have time to grow. Figure \ref{fig:shock_str} shows the
structure of the cloud ({\it left panel}) and the dependence of the
physical properties of the cloud on the Mach number of the inflowing
streams ({\it right panel}).

\articlefiguretwo{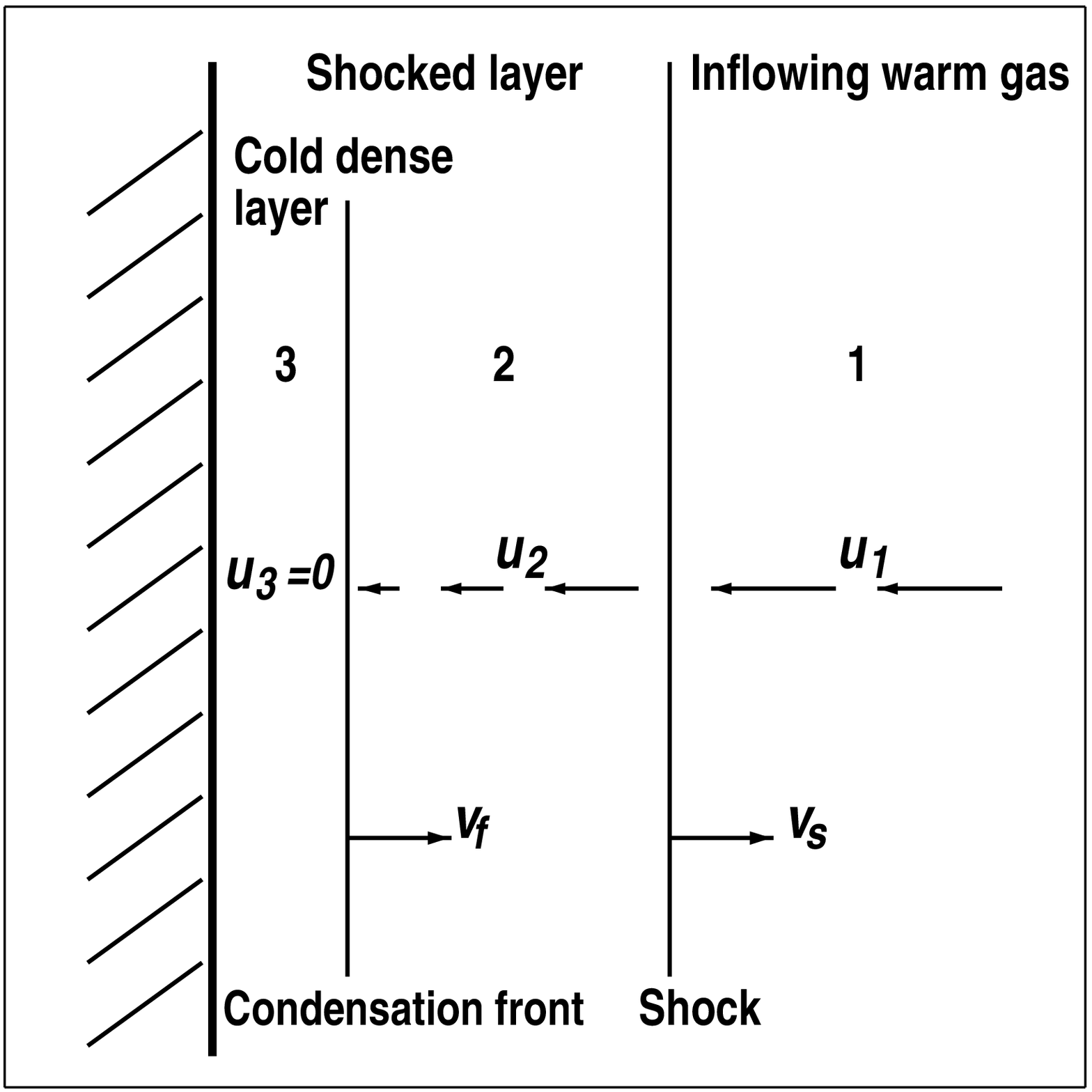}{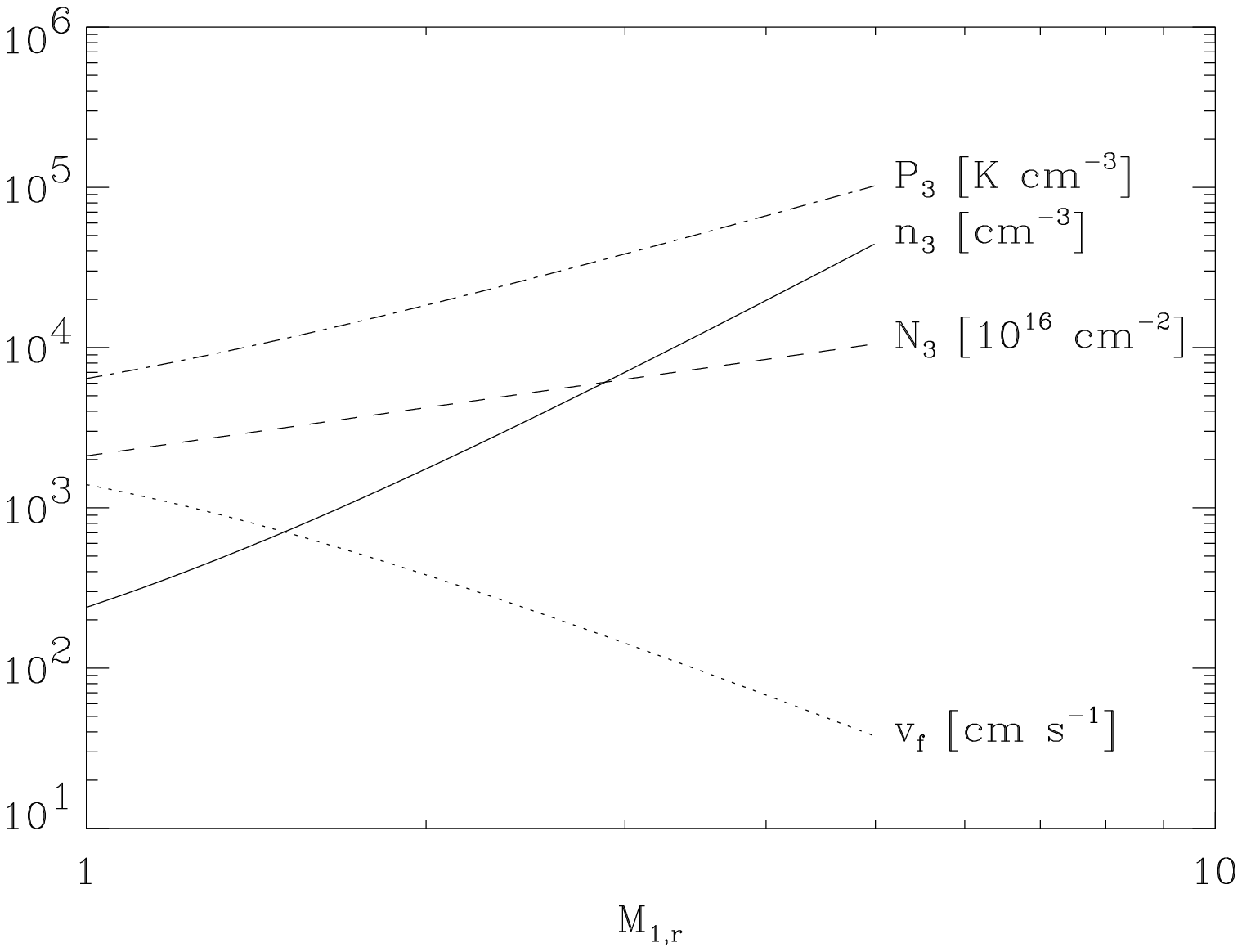}
{fig:shock_str} {{\it Left panel:} Schematic illustration of the structure of a
cloud formed by the collision of oppositely-directed WNM streams. The
streams are assumed to collide along the horizontal axis. Only the right
half of the system is shown, with the left stream replaced by a
wall. The inflowing gas approaches from the right, and shocks at the
interface between regions 1 and 2. The shock heating throws the
post-shock gas out of thermal balance between heating and cooling, and
so the gas cools as it flows towards the 2-3 interface. At the latter, the
gas undergoes a phase transition (condenses) to the CNM, and settles
into a dense, cold layer (region 3). The phase transition front moves
outwards from the collision wall at speed $v_{\rm f}$. {\it Right:}
Pressure $P$, number density $n$, outwards speed of the phase transition
front $v_{\rm f}$, and column density $N$ after 1 Myr of evolution, of
the thin layer, all as a funtion of the Mach number of the inflow
stream. From \cite{2006ApJ...643..245V}.}

One important implication of the study by \citet{2006ApJ...643..245V} is
that, in its initial stages, the forming cloud is expected to be a thin
CNM sheet, since it forms at the essentially two-dimensional interface
between the colliding streams.  After a few Myr of evolution, the
predicted thin sheet has column densities that agree very well with the
CNM sheets reported by \citet{2003ApJ...586.1067H}. Thus, it is
suggested that such thin CNM sheets may constitute the earliest phases
of MC evolution. Note, however, that a GMC may never form if the
mass involved in the streaming flows is not high enough to attain the
column densities necessary for molecule formation.

\section{Maturity: Gravitational contraction and star formation (child
bearing)} \label{sec:onset_SF} 

\subsection{A distribution of collapse timescales} \label{sec:tff_distr}

As discussed in \S \ref{sec:M2FR}, as a cloud grows, it should become
molecular, supercritical, and gravitationally unstable at roughly the
same time. This result implies that by the time a GMC forms, it should
be contracting gravitationally, since it cannot be supported by the
magnetic field because it is already supercritical, and it is not
forming stars yet, so no additional turbulence that can support the
cloud can be injected into it yet.

However, the initial turbulence produced by the convergent gas streams
has a crucial role in the subsequent development of the cloud and its SF
activity. Numerical simulations by various groups
\citep{2002ApJ...564L..97K, 2005A&A...433....1A, 2005ApJ...633L.113H,
2006ApJ...643..245V} have shown that the ``clouds'' are actually a
mixture of warm and cold gas \citep[see
also][]{2006ApJ...647..404H}. This is illustrated in Fig.\
\ref{fig:MC_struc} ({\it left panel}), which shows the granular, fractal
density structure observed in a high-resolution ($10000^2$ zones)
numerical simulation by \citet{2007A&A...465..431H}. The density field
has a wide probability density distribution function PDF (Fig.\
\ref{fig:MC_struc}, {\it right panel}), and thus, when gravity is
included, this density PDF implies the existence of a wide distribution
of free-fall times \citep[Fig.\
\ref{fig:tff_distr};][]{2008ApJ...689..290H}, with the bulk of the mass
remaining at relatively low densities . This implies that a small
fraction of the mass in the cloud will collapse, and thus SF will begin,
{\it before} the global collapse of cloud as a whole is completed, on
timescales $\sim 10$ Myr.

\articlefiguretwo{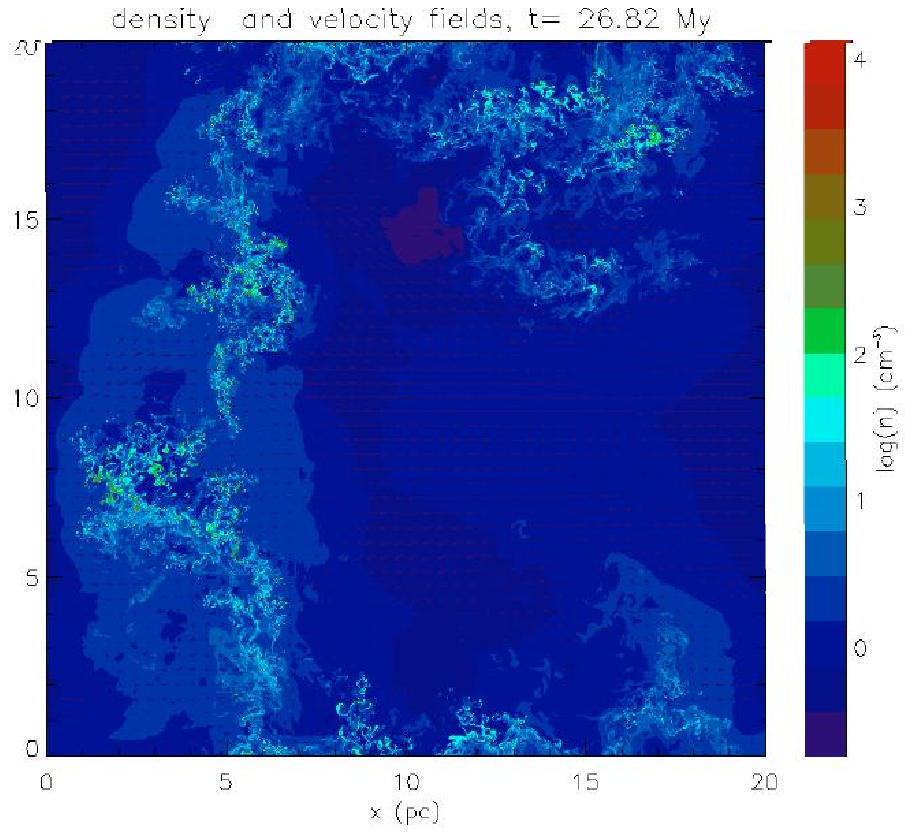}{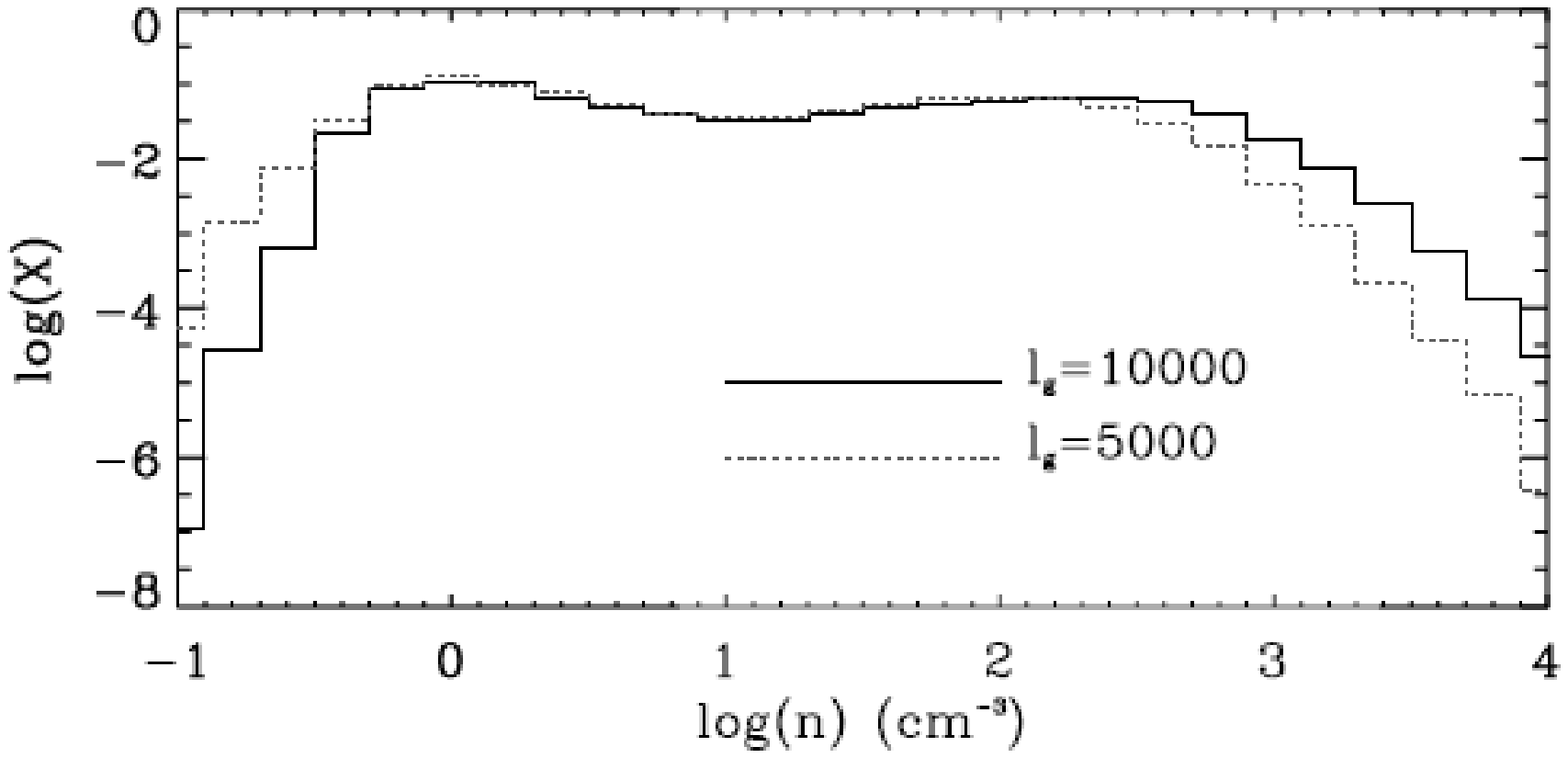}{fig:MC_struc}
{{\it Left:} Density and velocity fields in a two-dimensional numerical
simulation of colliding flows in a 20-pc numerical box, with a
resolution down to a scale of $2 \times 10^{-3}$ pc. Note the extremely
granular texture of the density field. {\it Right:} Volume-weighted
probability density distribution (PDF) of the density field for this
simulation ({\it solid line}) and for a similar one with lower (half)
the resolution ({\it dotted line}). From \citet{2007A&A...465..431H}. }

\articlefigure{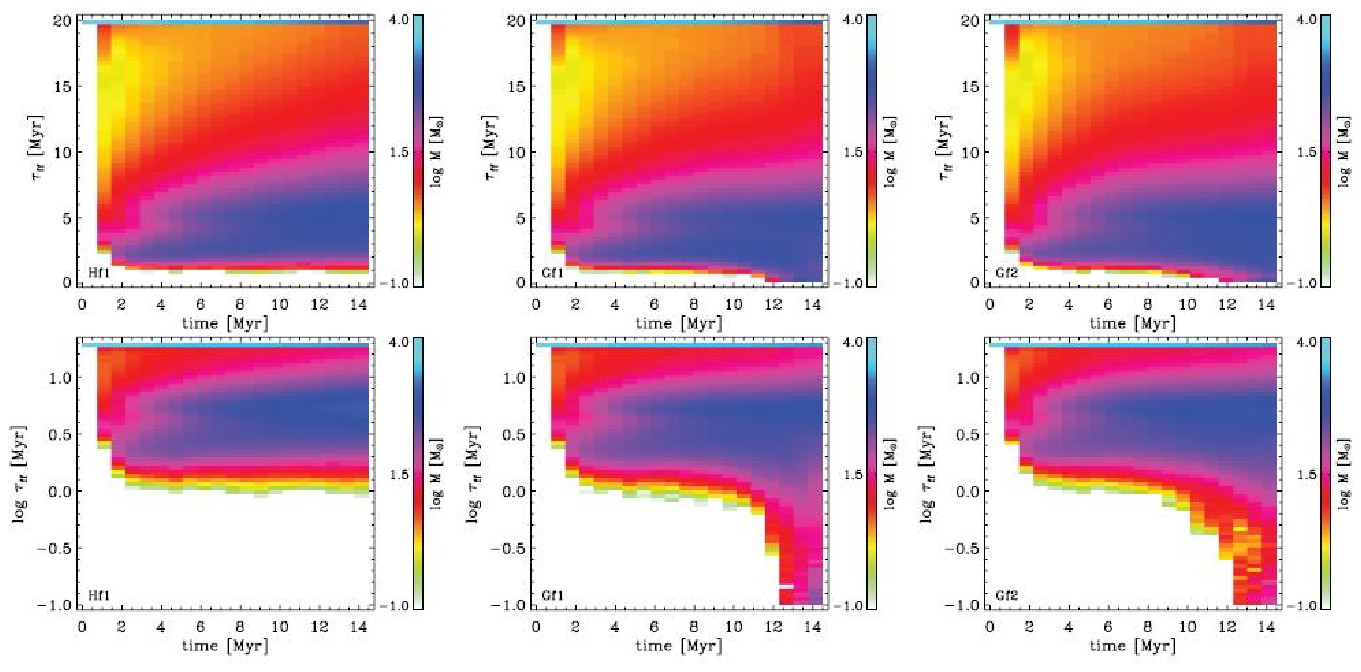}{fig:tff_distr}{Mass at a given
free-fall time as a funtion of time in three different numerical
simulations of cloud formation by colliding streams by
\citet{2008ApJ...689..290H}, with the free-fall time axis binned
linearly in the {\it top} panels, and logarithmically in the {\it
bottom} ones. It is seen that the bulk of the mass remains at low
densities, with long free-fall times. From \citet{2008ApJ...689..290H}.}

This kind of evolution, which we refer to as {\it hierarchical
gravitational collapse} was first observed numerically in the
simulations by \citet{2007ApJ...657..870V}, and is very similar to the
notion of hierarchical gravitational fragmentation introduced long ago
by \citet{1953ApJ...118..513H}, except that in that proposal, the
density fluctuations were assumed to be linear, and so they all had
essentially the same free-fall time as the large-scale cloud, while here
they are nonlinear, with shorter timescales than the cloud, due to the
initial turbulence induced by the very process of assembly of the
cloud.


\subsection{Low- and high-mass star-forming regions}
\label{sec:SF_masses}

The distribution of density fluctuations produced by the initial
turbulence has another implication: a clump mass spectrum is produced.
From high-resolution simulations, \citet{2007A&A...465..431H} reported a
power-law shape of the spectrum at high masses of the form $d N/dM
\propto M^{-\beta}$, with $\beta \sim 1.7$, in good agreement with
observational determinations of the mass spectrum for CO clumps
\citep{1998A&A...336..150M, 1998A&A...329..249K,
1998A&A...331L..65H}. It is well known that this form of the mass
spectrum implies that most of the mass is in the more massive clumps,
even though they are less numerous.

Of course, the less massive structures are expected to be embedded in the
larger-mass, lower-density ones in a hierarchical fashion
\citep{1994ApJ...423..681V}. This implies that the low-mass structures
should be expected to terminate their collapse and form stars earlier
than the more massive ones, and so a prediction from the present
scenario is that a large molecular complex should contain a relatively
large number of low-mass, somewhat older (by a few Myr) star-forming
regions, and a smaller number of massive regions, formed later, yet
containing most of the mass. The latter is consistent with the well
known result that most stars are formed in massive, cluster-forming
regions \citep{2003ARA&A..41...57L}. Of course, this picture
neglects triggering of secondary SF by previous generations of stars,
which complicates the picture.

This scenario has been quantified in a numerical simulation by
\citet{2009ApJ...707.1023V}, who selected a typical example of the
early-forming, low-mass regions, and the most massive cloud formed by
collapse of the cloud complex at large from a numerical simulation of
cloud formation by \citet{2007ApJ...657..870V}. This simulation used a
similar setup to that of the simulations by \citet{2009MNRAS.398.1082B},
discussed in \S \ref{sec:M2FR}, but with no magnetic
fields. \citet{2009ApJ...707.1023V} showed that indeed the masses,
sizes, velocity dispersions, and SFRs of the two regions were
respectively consistent with observations of such types of regions.
Moreover, they compared the distributions of masses, sizes and densities
of the dense cores within the massive region, showing that they were
very similar to the corresponding distributions for the set of cores in
the Cygnus-X region by \citet{2007A&A...476.1243M}. In summary, this
analysis suggested that the formation of low- and high-mass star-forming
regions by this mechanism is viable.

\section{Old age: Stellar feedback (popping all over)}
\label{sec:feedback}

Once SF begins in a cloud, the newly formed stars feedback on it via
either low-mass star outflows, which inject momentum, or ionizing
radiation and supernova explosions from massive stars. The effect of
this feedback on the parent cloud has been extensively studied by
numerous groups. For a detailed discussion, see the review by
V\'azquez-Semadeni (2011, in preparation). Here we focus only on whether
the feedback is able to bring the parent cloud to a quasi-static
equilibrium or whether, instead, it ends up dispersing the cloud. This
is an issue about which much effort is currently being devoted, and
because of that, no conclusive answer is yet available.

In recent years, 
simulations of multiple jets in parsec-scale clumps
\citep{2007ApJ...662..395N, 2009ApJ...695.1376C} have suggested that 
bipolar outflows are sufficient to drive and maintain the turbulence in
parsec-scale clumps, and to maintain the latter in a near-hydrostatic
equilibrium.  However, in those works the clumps occupied the entire
numerical box, and therefore the simulations lacked the contracting
motions of the rest of the MC in which they are embedded.

A simplified model of the effect of massive-star feedback through
HII-region expansion on the {\it global} evolution of GMCs was performed by
\citet{2006ApJ...653..361K}. These authors computed the time-dependent
virial {\it balance} (not necessarily equilibrium) of a spherical GMC
under the effect of its self-gravity and the energy injection of its
embedded HII regions. In their simplified model, the SFR self-regulates,
and causes oscillations of the clouds, which are finally dispersed after
a few crossing times.

Full numerical simulations of HII-region feedback in the context of a
fully-evolving and contracting GMC have been recently performed by
\citet{2010ApJ...715.1302V}, who again used the setup described in \S
\ref{sec:M2FR}, now in adaptive mesh refinement (AMR) simulations which
allowed an effective resolution of $\sim 0.03$ pc, and that included
feedback from a single-mass population of stars. These authors found
that the feedback affects the immediate surroundings of the recently
formed stars, thus reducing the SFR, but is not capable of reverting the
global contraction of the GMC. Similar results have been obtained in
high-resolution simulations at the clump-scale with outflow feedback by
\citet{2010ApJ...709...27W}. Those authors have found that the accretion
that feeds the most massive star is not restricted to the core in which
the star is being formed directly, but rather can be traced out to the
scale of the whole clump containing the core, in spite of the fact that
the outflows distort this flow, reducing the SFE of the system.

In summary, the available numerical evidence suggests that the scenario
of gravitationally contracting MCs is maintained even in the presence of
stellar feedback. The latter may be an important source of energy for
driving the turbulence at the clump (parsec) scale, but has a harder
time halting the gravitational contraction at the scale of the whole GMC
(tens of parsecs). The calculations by \citet{2006ApJ...653..361K} and
\citet{2010ApJ...715.1302V} suggest that low-mass clouds are more
readily destroyed by the feedback than more massive ones. Nevertheless,
the numerical experiments performed so far are relatively scarce, and
have used a limited set of initial conditions. A more complete coverage
of parameter space, such as that conducted by
\citet{2010MNRAS.406.1875R} for the variability of the SFR with the
initial conditions, and with a more accurate modeling of the stellar
feedback, is necessary to better understand the details of cloud dispersal.

\section{Conclusions} \label{sec:conclusions}

\subsection{Summary} \label{sec:summary}

The scenario of MC formation and evolution (under solar neighborhood
conditions) discussed in the previous sections can be summarized as
follows:

\begin{itemize}

\item The route to the formation of a GMC starts with a large-scale,
moderately supersonic converging motion in the WNM, which may be driven
either by large-scale instabilities, the passage of a spiral-arm or
supernova shock, or by intermediate-scale generic turbulence in
the WNM.

\item The compression nonlinearly triggers a phase transition to the
CNM, forming a large, though not very dense, cold atomic cloud. The
earliest stages of these clouds may constitute the thin CNM clouds
reported recently by \citet{2003ApJ...586.1067H}. At later times, a
combination of the Kelvin-Helmholz, Rayleigh-Taylor, nonlinear
thin-shell and thermal instabilities destabilizes the cloud, rendering
it turbulent and clumpy.

\item If sufficient mass is available in the converging flow that a
column density $\sim 1.5 \times 10^{21} \psc$ is reached in the newly
formed cloud, then the cloud begins to be dominated by self-gravity
(rather than by the confining pressure [thermal+ram] of its
surroundings) and also reaches a high enough extinction ($A_V \sim 1$)
to allow the formation of CO molecules. Moreover, for a fiducial value
of the mean Galactic magnetic field of $5 \mu$G, such a cloud should be
near to becoming magnetically supercritical as well. So, for solar
neighborhood conditions, a growing cloud should become molecular,
self-gravitating and magnetically supercritical at approximately the
same time.

\item A cloud that has reached such a column density then begins to
contract gravitationally. However, the clumps produced by the various
instabilities, which have shorter free-fall times than the bulk of the
cloud, culminate their collapses before the bulk of the cloud does. Star
formation thus begins a few Myr after the cloud's global contraction has
started, but a few Myr before the bulk of the cloud cuminates its own
collapse.

\item The collapse of isolated, low-mass clumps produces scattered
low-mas star-form\-ing regions, while the collapse of the bulk of the
cloud produces high-mass star-form\-ing regions.

\item The termination of a SF episode in a cloud is still not fully
understood from the avilable numerical simulations. More complete
coverage of parameter space and with more detailed modeling of the
feedback is necessary. In any case, the attainment of a nearly
hydrostatic equilibrium appears very difficult. Instead, it appears that
a cloud may begin to be destroyed or dispersed locally, while the outer
layers may still be falling in, establishing a complicated flow with
both infall and outflow.

\end{itemize}

\subsection{Implications: hierarchical gravitational contraction}
\label{sec:implications} 

The evolutionary scenario for MCs described here strongly suggests the
ubiquitous existence of generalized gravitational contraction in MCs and
their substructure, given the approximate simultaneity of the onset of
gravitational contraction with the onset of molecule formation and the
attainement of magnetic supercriticality.
This implication is in fact consistent with a growing body
of observational results. In particular, principal component analysis of
the contributions to the velocity dispersion in MCs and their
substructure invariably show a ``dipolar'' main component, indicating
that the dominant contribution comes from a large-scale (i.e., at the
scale of the full structure observed) velocity gradient
\citep{2007IAUS..237....9H, 2009A&A...504..883B} which is consistent
with a global contraction or shear of the cloud, but inconsistent with
solid-body rotation (M. Heyer, priv. comm., 2010). Also, studies
comparing specific regions with numerical simulations
\citep{2007ApJ...654..988H, 2007A&A...464..983P} have shown that those
regions are well modeled by gravitationally contracting structures.
Finally, recent studies of massive star-forming regions by
\citep{2009ApJ...706.1036G, 2010arXiv1009.0598C} have provided evidence
of the existence of {\it hierarchical accretion flows} at multiple
scales in the regions. The notion of hierarchical gravitational
fragmentation has recently been formulated analytically by
\citet{2008MNRAS.385..181F}, who have proposed the existence of a
gravitationally driven cascade from the large to the small scales in
MCs, analogous to a turbulent cascade, except that the quantity being
cascaded is mass rather than energy. 

Moreover, recent observations by
\citet{2009ApJ...699.1092H} having much higher angular and spectral
resolution and
higher dynamic range in column density than earlier studies
\citep[e.g.,][]{1987ApJ...319..730S}, suggest that MCs do not, after all,
have a roughly constant column density $\Sigma$, but rather span up to
two orders of magnitude in this variable, and that the velocity
dispersion actually scales with size {\it and} column density as 
\begin{equation}
\sigma_v \propto (\Sigma L)^{1/2}.
\label{eq:Heyer_rel}
\end{equation}
Recently, \citet{2010arXiv1009.1583B} have proposed that this scaling is
exactly what is expected from a hierarchical gravitational cascade, in
which, rather than virial {\it equilibrium}, the governing relation is
simply energy conservation during the contraction, so that the
gravitational and kinetic energies satisfy 
\begin{equation}
|E_{\rm g}| = E_{\rm k},
\label{eq:ener_cons}
\end{equation}
from which eq.\ (\ref{eq:Heyer_rel}) follows directly. It is noteworthy
that, at face value, eq.\ (\ref{eq:ener_cons}) seems to fit the data
better than the virial equilibrium condition $|E_{\rm g}| = 2 E_{\rm
k}$, as shown in Fig.\ \ref{fig:Heyer_law}, although large
uncertainties, especially in the mass determinations, prevent any firm
conclusions. Interestingly, note that free-fall implies {\it larger}
velocities than virial equilibrium, contrary to the standard notion that
velocities higher than virial imply lack of gravitational binding.

\articlefigure{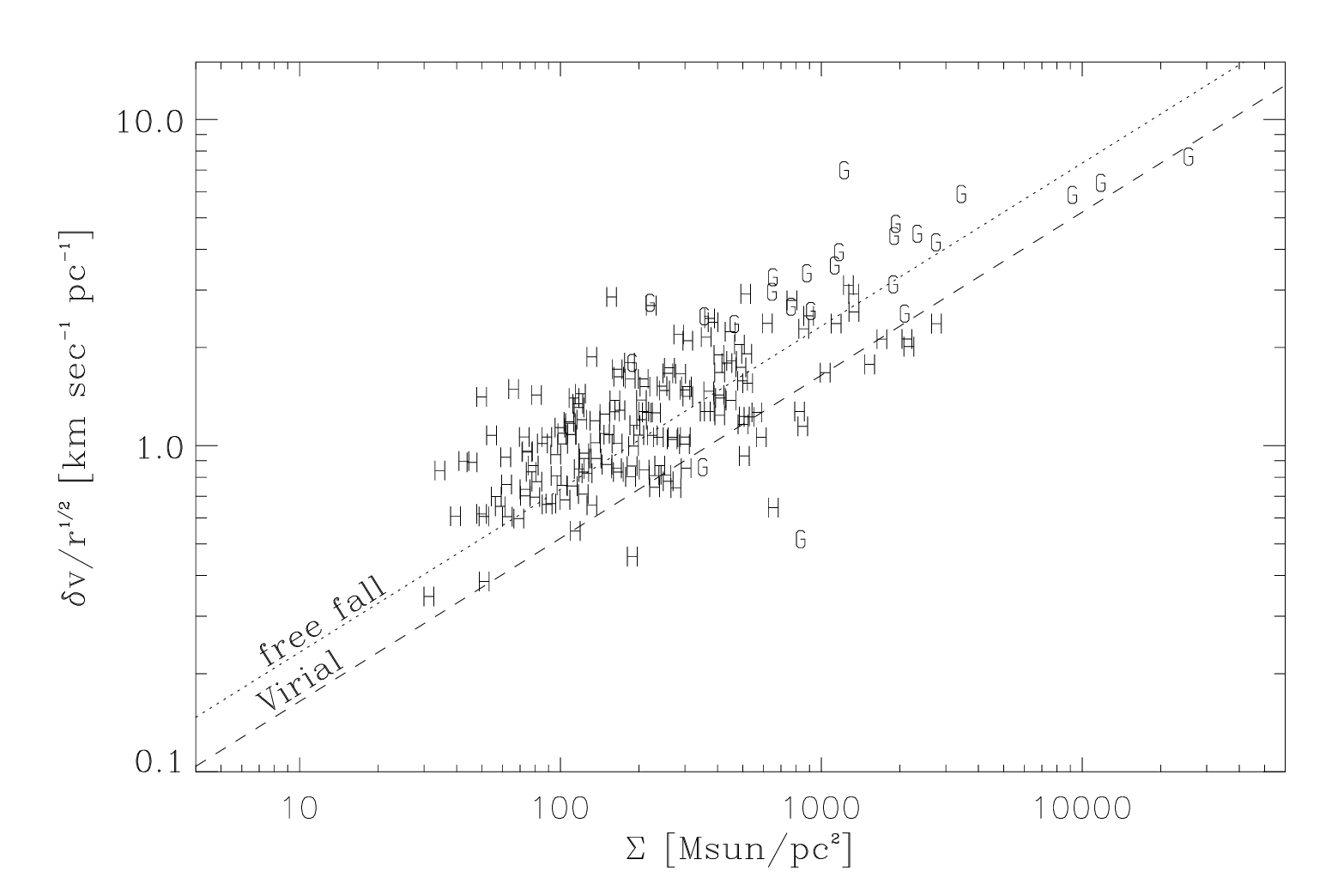}{fig:Heyer_law}{Massive dense cores
from \citet[][labeled ``G'']{2009ApJ...705..123G}, and clouds and
clumps from \citet[][labeled ``H'']{2009ApJ...699.1092H} in the
$\sigma_v/R^{1/2}$ vs.\ $\Sigma$ plane, where $\sigma_v$ is the velocity
dispersion, $R$ is the region size, and $\Sigma$ is the mass column
density. The straight lines show the loci of virial equilibrium,
$|E_{\rm g}| = 2 E_{\rm k}$, and of energy conservation under free-fall,
$|E_{\rm g}| = E_{\rm k}$. From \citet{2010arXiv1009.1583B}.}

It is important to remark that the hypothesis that MCs might be undergoing
generalized gravitational contraction is not new. It was first proposed
by \citet{1974ApJ...189..441G}. However, it was soon deemed untenable by
\citet{1974ARA&A..12..279Z} who argued that, if the clouds were in
free-fall, then a simple estimate obtained by dividing the total
molecular mass in the Galaxy by the mean free-fall time would imply a
SFR roughly two orders of magnitude larger than observed. However, this
conundrum may not pose a serious problem if the star-forming activity of
the clouds is terminated prematurely by the feedback from the first
stars formed. More work to understand the details of this process is
still needed.

\subsection{Final remarks} \label{sec:final_rem}

The picture of MC evolution described in this review appears promising
as a self-consistent scenario that connects the dynamics of the ISM at
large with the physical properties of MCs and their star-forming
properties. However, several features still need to be worked out in
more detail, such as the precise form in which stellar feedback
terminates a SF episode and disperses a cloud locally, and at what
scales can this process be considered in an averaged sense, if at all. 
Also, more quantitative statistical issues, such as the fraction of the
time, and of the stellar production of the clouds, is spent under the
magnetically subcritical and supercritical regimes, respectively.

Finally, it is important to remark that the present scenario is likely to
only apply directly to solar neighborhood-like conditions. In
particular, it may need modification for regions like the molecular ring
of the Galaxy, or to galaxies where the disk is mostly molecular, as in
those cases the atomic component may be absent, at least locally. In
those cases, it is possible that the entire molecular ring or disk is
the equivalent of the isolated GMCs we have discussed here, and that the
phase transition occurs at the boundaries of these regions, both
radially and vertically. More work is needed in order to assess this.

\acknowledgements 

I warmfully thank all the colleagues who have helped shaping my ideas
about the subject of this review over the years, especially Javier
Ballesteros-Paredes, Lee Hartmann, Patrick Hennebelle, and Mordecai Mac
Low.  Gilberto G\'omez and Pedro Col\'\i n have provided invaluable
expertise in translating those ideas into numerical simulations. This
work has been funded in part by CONACYT grant 102488 to E.V.-S.

\bibliography{aspauthor}

\begin{thebibliography}{}
\expandafter\ifx\csname natexlab\endcsname\relax\def\natexlab#1{#1}\fi
\expandafter\ifx\csname url\endcsname\relax
  \def\url#1{\texttt{#1}}\fi
\expandafter\ifx\csname urlprefix\endcsname\relax\def\urlprefix{URL }\fi
\providecommand{\eprint}[2][]{\url{#2}}

\bibitem[{{Audit} \& {Hennebelle}(2005)}]{2005A&A...433....1A}
{Audit}, E., \& {Hennebelle}, P. 2005, \aap, 433, 1.
  \eprint{arXiv:astro-ph/0410062}

\bibitem[{{Ballesteros-Paredes}
  et~al.(1999{\natexlab{a}}){Ballesteros-Paredes}, {Hartmann}, \&
  {V{\'a}zquez-Semadeni}}]{1999ApJ...527..285B}
{Ballesteros-Paredes}, J., {Hartmann}, L., \& {V{\'a}zquez-Semadeni}, E.
  1999{\natexlab{a}}, \apj, 527, 285. \eprint{arXiv:astro-ph/9907053}

\bibitem[{{Ballesteros-Paredes} et~al.(2010){Ballesteros-Paredes}, {Hartmann},
  {V{\'a}zquez-Semadeni}, {Heitsch}, \&
  {Zamora-Avil{\'e}s}}]{2010arXiv1009.1583B}
{Ballesteros-Paredes}, J., {Hartmann}, L.~W., {V{\'a}zquez-Semadeni}, E.,
  {Heitsch}, F., \& {Zamora-Avil{\'e}s}, M.~A. 2010, ArXiv e-prints.
  \eprint{1009.1583}

\bibitem[{{Ballesteros-Paredes} et~al.(2007){Ballesteros-Paredes}, {Klessen},
  {Mac Low}, \& {V\'azquez-Semadeni}}]{2007prpl.conf...63B}
{Ballesteros-Paredes}, J., {Klessen}, R.~S., {Mac Low}, M., \&
  {V\'azquez-Semadeni}, E. 2007, Protostars and Planets V, 63.
  \eprint{arXiv:astro-ph/0603357}

\bibitem[{{Ballesteros-Paredes}
  et~al.(1999{\natexlab{b}}){Ballesteros-Paredes}, {V{\'a}zquez-Semadeni}, \&
  {Scalo}}]{1999ApJ...515..286B}
{Ballesteros-Paredes}, J., {V{\'a}zquez-Semadeni}, E., \& {Scalo}, J.
  1999{\natexlab{b}}, \apj, 515, 286. \eprint{arXiv:astro-ph/9806059}

\bibitem[{{Banerjee} et~al.(2009){Banerjee}, {V{\'a}zquez-Semadeni},
  {Hennebelle}, \& {Klessen}}]{2009MNRAS.398.1082B}
{Banerjee}, R., {V{\'a}zquez-Semadeni}, E., {Hennebelle}, P., \& {Klessen},
  R.~S. 2009, \mnras, 398, 1082. \eprint{0808.0986}

\bibitem[{{Bania} \& {Lyon}(1980)}]{1980ApJ...239..173B}
{Bania}, T.~M., \& {Lyon}, J.~G. 1980, \apj, 239, 173

\bibitem[{{Blitz}(1993)}]{1993prpl.conf..125B}
{Blitz}, L. 1993, in Protostars and Planets III, edited by {E.~H.~Levy \&
  J.~I.~Lunine}, 125

\bibitem[{{Blitz} \& {Shu}(1980)}]{1980ApJ...238..148B}
{Blitz}, L., \& {Shu}, F.~H. 1980, \apj, 238, 148

\bibitem[{{Bourke} et~al.(2001){Bourke}, {Myers}, {Robinson}, \&
  {Hyland}}]{2001ApJ...554..916B}
{Bourke}, T.~L., {Myers}, P.~C., {Robinson}, G., \& {Hyland}, A.~R. 2001, \apj,
  554, 916. \eprint{arXiv:astro-ph/0102469}

\bibitem[{{Brunt} et~al.(2009){Brunt}, {Heyer}, \& {Mac
  Low}}]{2009A&A...504..883B}
{Brunt}, C.~M., {Heyer}, M.~H., \& {Mac Low}, M. 2009, \aap, 504, 883.
  \eprint{0910.0398}

\bibitem[{{Carroll} et~al.(2009){Carroll}, {Frank}, {Blackman}, {Cunningham},
  \& {Quillen}}]{2009ApJ...695.1376C}
{Carroll}, J.~J., {Frank}, A., {Blackman}, E.~G., {Cunningham}, A.~J., \&
  {Quillen}, A.~C. 2009, \apj, 695, 1376. \eprint{0805.4645}

\bibitem[{{Chiang} \& {Bregman}(1988)}]{1988ApJ...328..427C}
{Chiang}, W., \& {Bregman}, J.~N. 1988, \apj, 328, 427

\bibitem[{{Csengeri} et~al.(2010){Csengeri}, {Bontemps}, {Schneider}, {Motte},
  \& {Dib}}]{2010arXiv1009.0598C}
{Csengeri}, T., {Bontemps}, S., {Schneider}, N., {Motte}, F., \& {Dib}, S.
  2010, ArXiv e-prints. \eprint{1009.0598}

\bibitem[{{de Avillez}(2000)}]{2000MNRAS.315..479D}
{de Avillez}, M.~A. 2000, \mnras, 315, 479. \eprint{arXiv:astro-ph/0001315}

\bibitem[{{de Avillez} \& {Breitschwerdt}(2004)}]{2004A&A...425..899D}
{de Avillez}, M.~A., \& {Breitschwerdt}, D. 2004, \aap, 425, 899.
  \eprint{arXiv:astro-ph/0407034}

\bibitem[{{de Avillez} \& {Breitschwerdt}(2005)}]{2005A&A...436..585D}
--- 2005, \aap, 436, 585. \eprint{arXiv:astro-ph/0502327}

\bibitem[{{Elmegreen}(1994)}]{1994ApJ...433...39E}
{Elmegreen}, B.~G. 1994, \apj, 433, 39

\bibitem[{{Engargiola} et~al.(2003){Engargiola}, {Plambeck}, {Rosolowsky}, \&
  {Blitz}}]{2003ApJS..149..343E}
{Engargiola}, G., {Plambeck}, R.~L., {Rosolowsky}, E., \& {Blitz}, L. 2003,
  \apjs, 149, 343. \eprint{arXiv:astro-ph/0308388}

\bibitem[{{Field}(1965)}]{1965ApJ...142..531F}
{Field}, G.~B. 1965, \apj, 142, 531

\bibitem[{{Field} et~al.(2008){Field}, {Blackman}, \&
  {Keto}}]{2008MNRAS.385..181F}
{Field}, G.~B., {Blackman}, E.~G., \& {Keto}, E.~R. 2008, \mnras, 385, 181

\bibitem[{{Field} et~al.(1969){Field}, {Goldsmith}, \&
  {Habing}}]{1969ApJ...155L.149F}
{Field}, G.~B., {Goldsmith}, D.~W., \& {Habing}, H.~J. 1969, \apjl, 155, L149+

\bibitem[{{Field} \& {Saslaw}(1965)}]{1965ApJ...142..568F}
{Field}, G.~B., \& {Saslaw}, W.~C. 1965, \apj, 142, 568

\bibitem[{{Franco} \& {Cox}(1986)}]{1986PASP...98.1076F}
{Franco}, J., \& {Cox}, D.~P. 1986, \pasp, 98, 1076

\bibitem[{{Galv{\'a}n-Madrid} et~al.(2009){Galv{\'a}n-Madrid}, {Keto}, {Zhang},
  {Kurtz}, {Rodr{\'{\i}}guez}, \& {Ho}}]{2009ApJ...706.1036G}
{Galv{\'a}n-Madrid}, R., {Keto}, E., {Zhang}, Q., {Kurtz}, S.,
  {Rodr{\'{\i}}guez}, L.~F., \& {Ho}, P.~T.~P. 2009, \apj, 706, 1036.
  \eprint{0910.2270}

\bibitem[{{Gibson} et~al.(2009){Gibson}, {Plume}, {Bergin}, {Ragan}, \&
  {Evans}}]{2009ApJ...705..123G}
{Gibson}, D., {Plume}, R., {Bergin}, E., {Ragan}, S., \& {Evans}, N. 2009,
  \apj, 705, 123. \eprint{0908.2643}

\bibitem[{{Goldreich} \& {Kwan}(1974)}]{1974ApJ...189..441G}
{Goldreich}, P., \& {Kwan}, J. 1974, \apj, 189, 441

\bibitem[{{Hartmann} et~al.(2001){Hartmann}, {Ballesteros-Paredes}, \&
  {Bergin}}]{2001ApJ...562..852H}
{Hartmann}, L., {Ballesteros-Paredes}, J., \& {Bergin}, E.~A. 2001, \apj, 562,
  852. \eprint{arXiv:astro-ph/0108023}

\bibitem[{{Hartmann} \& {Burkert}(2007)}]{2007ApJ...654..988H}
{Hartmann}, L., \& {Burkert}, A. 2007, \apj, 654, 988.
  \eprint{arXiv:astro-ph/0609679}

\bibitem[{{Heiles} \& {Troland}(2003)}]{2003ApJ...586.1067H}
{Heiles}, C., \& {Troland}, T.~H. 2003, \apj, 586, 1067.
  \eprint{arXiv:astro-ph/0207105}

\bibitem[{{Heiles} \& {Troland}(2005)}]{2005ApJ...624..773H}
--- 2005, \apj, 624, 773. \eprint{arXiv:astro-ph/0501482}

\bibitem[{{Heithausen} et~al.(1998){Heithausen}, {Bensch}, {Stutzki},
  {Falgarone}, \& {Panis}}]{1998A&A...331L..65H}
{Heithausen}, A., {Bensch}, F., {Stutzki}, J., {Falgarone}, E., \& {Panis},
  J.~F. 1998, \aap, 331, L65

\bibitem[{{Heitsch} et~al.(2005){Heitsch}, {Burkert}, {Hartmann}, {Slyz}, \&
  {Devriendt}}]{2005ApJ...633L.113H}
{Heitsch}, F., {Burkert}, A., {Hartmann}, L.~W., {Slyz}, A.~D., \& {Devriendt},
  J.~E.~G. 2005, \apjl, 633, L113. \eprint{arXiv:astro-ph/0507567}

\bibitem[{{Heitsch} \& {Hartmann}(2008)}]{2008ApJ...689..290H}
{Heitsch}, F., \& {Hartmann}, L. 2008, \apj, 689, 290. \eprint{0808.1078}

\bibitem[{{Heitsch} et~al.(2006){Heitsch}, {Slyz}, {Devriendt}, {Hartmann}, \&
  {Burkert}}]{2006ApJ...648.1052H}
{Heitsch}, F., {Slyz}, A.~D., {Devriendt}, J.~E.~G., {Hartmann}, L.~W., \&
  {Burkert}, A. 2006, \apj, 648, 1052. \eprint{arXiv:astro-ph/0605435}

\bibitem[{{Hennebelle} \& {Audit}(2007)}]{2007A&A...465..431H}
{Hennebelle}, P., \& {Audit}, E. 2007, \aap, 465, 431

\bibitem[{{Hennebelle} \& {Inutsuka}(2006)}]{2006ApJ...647..404H}
{Hennebelle}, P., \& {Inutsuka}, S. 2006, \apj, 647, 404.
  \eprint{arXiv:astro-ph/0510389}

\bibitem[{{Hennebelle} \& {P{\'e}rault}(1999)}]{1999A&A...351..309H}
{Hennebelle}, P., \& {P{\'e}rault}, M. 1999, \aap, 351, 309

\bibitem[{{Hennebelle} \& {P{\'e}rault}(2000)}]{2000A&A...359.1124H}
--- 2000, \aap, 359, 1124

\bibitem[{{Heyer} et~al.(2009){Heyer}, {Krawczyk}, {Duval}, \&
  {Jackson}}]{2009ApJ...699.1092H}
{Heyer}, M., {Krawczyk}, C., {Duval}, J., \& {Jackson}, J.~M. 2009, \apj, 699,
  1092. \eprint{0809.1397}

\bibitem[{{Heyer} \& {Brunt}(2007)}]{2007IAUS..237....9H}
{Heyer}, M.~H., \& {Brunt}, C. 2007, in IAU Symposium, edited by
  {B.~G.~Elmegreen \& J.~Palous}, vol. 237 of IAU Symposium, 9.
  \eprint{arXiv:astro-ph/0611492}

\bibitem[{{Hoyle}(1953)}]{1953ApJ...118..513H}
{Hoyle}, F. 1953, \apj, 118, 513

\bibitem[{{Hunter} et~al.(1986){Hunter}, {Sandford}, {Whitaker}, \&
  {Klein}}]{1986ApJ...305..309H}
{Hunter}, J.~H., Jr., {Sandford}, M.~T., II, {Whitaker}, R.~W., \& {Klein},
  R.~I. 1986, \apj, 305, 309

\bibitem[{{Kawamura} et~al.(2009){Kawamura}, {Mizuno}, {Minamidani},
  {Filipovi{\'c}}, {Staveley-Smith}, {Kim}, {Mizuno}, {Onishi}, {Mizuno}, \&
  {Fukui}}]{2009ApJS..184....1K}
{Kawamura}, A., {Mizuno}, Y., {Minamidani}, T., {Filipovi{\'c}}, M.~D.,
  {Staveley-Smith}, L., {Kim}, S., {Mizuno}, N., {Onishi}, T., {Mizuno}, A., \&
  {Fukui}, Y. 2009, \apjs, 184, 1. \eprint{0908.1168}

\bibitem[{{Kim} et~al.(2002){Kim}, {Ostriker}, \&
  {Stone}}]{2002ApJ...581.1080K}
{Kim}, W., {Ostriker}, E.~C., \& {Stone}, J.~M. 2002, \apj, 581, 1080.
  \eprint{arXiv:astro-ph/0208414}

\bibitem[{{Korpi} et~al.(1999){Korpi}, {Brandenburg}, {Shukurov}, {Tuominen},
  \& {Nordlund}}]{1999ApJ...514L..99K}
{Korpi}, M.~J., {Brandenburg}, A., {Shukurov}, A., {Tuominen}, I., \&
  {Nordlund}, {\AA}. 1999, \apjl, 514, L99

\bibitem[{{Koyama} \& {Inutsuka}(2002)}]{2002ApJ...564L..97K}
{Koyama}, H., \& {Inutsuka}, S. 2002, \apjl, 564, L97.
  \eprint{arXiv:astro-ph/0112420}

\bibitem[{{Kramer} et~al.(1998){Kramer}, {Stutzki}, {Rohrig}, \&
  {Corneliussen}}]{1998A&A...329..249K}
{Kramer}, C., {Stutzki}, J., {Rohrig}, R., \& {Corneliussen}, U. 1998, \aap,
  329, 249

\bibitem[{{Krumholz} et~al.(2006){Krumholz}, {Matzner}, \&
  {McKee}}]{2006ApJ...653..361K}
{Krumholz}, M.~R., {Matzner}, C.~D., \& {McKee}, C.~F. 2006, \apj, 653, 361.
  \eprint{arXiv:astro-ph/0608471}

\bibitem[{{Kulkarni} \& {Heiles}(1987)}]{1987ASSL..134...87K}
{Kulkarni}, S.~R., \& {Heiles}, C. 1987, in Interstellar Processes, edited by
  {D.~J.~Hollenbach \& H.~A.~Thronson Jr.}, vol. 134 of Astrophysics and Space
  Science Library, 87

\bibitem[{{Kwan}(1979)}]{1979ApJ...229..567K}
{Kwan}, J. 1979, \apj, 229, 567

\bibitem[{{Lada} \& {Lada}(2003)}]{2003ARA&A..41...57L}
{Lada}, C.~J., \& {Lada}, E.~A. 2003, \araa, 41, 57.
  \eprint{arXiv:astro-ph/0301540}

\bibitem[{{Larson}(1981)}]{1981MNRAS.194..809L}
{Larson}, R.~B. 1981, \mnras, 194, 809

\bibitem[{{Mac Low} \& {Klessen}(2004)}]{2004RvMP...76..125M}
{Mac Low}, M., \& {Klessen}, R.~S. 2004, Reviews of Modern Physics, 76, 125.
  \eprint{arXiv:astro-ph/0301093}

\bibitem[{{McKee}(1989)}]{1989ApJ...345..782M}
{McKee}, C.~F. 1989, \apj, 345, 782

\bibitem[{{McKee} \& {Ostriker}(2007)}]{2007ARA&A..45..565M}
{McKee}, C.~F., \& {Ostriker}, E.~C. 2007, \araa, 45, 565. \eprint{0707.3514}

\bibitem[{{McKee} \& {Ostriker}(1977)}]{1977ApJ...218..148M}
{McKee}, C.~F., \& {Ostriker}, J.~P. 1977, \apj, 218, 148

\bibitem[{{Mestel}(1985)}]{1985prpl.conf..320M}
{Mestel}, L. 1985, in Protostars and Planets II, edited by {D.~C.~Black \&
  M.~S.~Matthews}, 320

\bibitem[{{Motte} et~al.(1998){Motte}, {Andre}, \&
  {Neri}}]{1998A&A...336..150M}
{Motte}, F., {Andre}, P., \& {Neri}, R. 1998, \aap, 336, 150

\bibitem[{{Motte} et~al.(2007){Motte}, {Bontemps}, {Schilke}, {Schneider},
  {Menten}, \& {Brogui{\`e}re}}]{2007A&A...476.1243M}
{Motte}, F., {Bontemps}, S., {Schilke}, P., {Schneider}, N., {Menten}, K.~M.,
  \& {Brogui{\`e}re}, D. 2007, \aap, 476, 1243. \eprint{0708.2774}

\bibitem[{{Mouschovias}(1991)}]{1991psfe.conf..449M}
{Mouschovias}, T.~C. 1991, in NATO ASIC Proc. 342: The Physics of Star
  Formation and Early Stellar Evolution, edited by {C.~J.~Lada \&
  N.~D.~Kylafis}, 449

\bibitem[{{Myers}(1978)}]{1978ApJ...225..380M}
{Myers}, P.~C. 1978, \apj, 225, 380

\bibitem[{{Myers} \& {Goodman}(1988)}]{1988ApJ...326L..27M}
{Myers}, P.~C., \& {Goodman}, A.~A. 1988, \apjl, 326, L27

\bibitem[{{Nakamura} \& {Li}(2007)}]{2007ApJ...662..395N}
{Nakamura}, F., \& {Li}, Z. 2007, \apj, 662, 395.
  \eprint{arXiv:astro-ph/0703152}

\bibitem[{{Nakano} \& {Nakamura}(1978)}]{1978PASJ...30..671N}
{Nakano}, T., \& {Nakamura}, T. 1978, \pasj, 30, 671

\bibitem[{{Norman} \& {Silk}(1980)}]{1980ApJ...238..158N}
{Norman}, C., \& {Silk}, J. 1980, \apj, 238, 158

\bibitem[{{Oort}(1954)}]{1954BAN....12..177O}
{Oort}, J.~H. 1954, Bull. Astron. Inst. Netherlands, 12, 177

\bibitem[{{Parker}(1966)}]{1966ApJ...145..811P}
{Parker}, E.~N. 1966, \apj, 145, 811

\bibitem[{{Passot} et~al.(1995){Passot}, {V\'azquez-Semadeni}, \&
  {Pouquet}}]{1995ApJ...455..536P}
{Passot}, T., {V\'azquez-Semadeni}, E., \& {Pouquet}, A. 1995, \apj, 455, 536.
  \eprint{arXiv:astro-ph/9601182}

\bibitem[{{Peretto} et~al.(2007){Peretto}, {Hennebelle}, \&
  {Andr{\'e}}}]{2007A&A...464..983P}
{Peretto}, N., {Hennebelle}, P., \& {Andr{\'e}}, P. 2007, \aap, 464, 983.
  \eprint{arXiv:astro-ph/0611277}

\bibitem[{{Pittard} et~al.(2005){Pittard}, {Dobson}, {Durisen}, {Dyson},
  {Hartquist}, \& {O'Brien}}]{2005A&A...438...11P}
{Pittard}, J.~M., {Dobson}, M.~S., {Durisen}, R.~H., {Dyson}, J.~E.,
  {Hartquist}, T.~W., \& {O'Brien}, J.~T. 2005, \aap, 438, 11.
  \eprint{arXiv:astro-ph/0504640}

\bibitem[{{Rosas-Guevara} et~al.(2010){Rosas-Guevara}, {V{\'a}zquez-Semadeni},
  {G{\'o}mez}, \& {Jappsen}}]{2010MNRAS.406.1875R}
{Rosas-Guevara}, Y., {V{\'a}zquez-Semadeni}, E., {G{\'o}mez}, G.~C., \&
  {Jappsen}, A. 2010, \mnras, 406, 1875. \eprint{0911.1795}

\bibitem[{{Rosen} \& {Bregman}(1995)}]{1995ApJ...440..634R}
{Rosen}, A., \& {Bregman}, J.~N. 1995, \apj, 440, 634

\bibitem[{{Rosen} et~al.(1993){Rosen}, {Bregman}, \&
  {Norman}}]{1993ApJ...413..137R}
{Rosen}, A., {Bregman}, J.~N., \& {Norman}, M.~L. 1993, \apj, 413, 137

\bibitem[{{Scoville} \& {Hersh}(1979)}]{1979ApJ...229..578S}
{Scoville}, N.~Z., \& {Hersh}, K. 1979, \apj, 229, 578

\bibitem[{{Shu}(1992)}]{1992phas.book.....S}
{Shu}, F.~H. 1992, {Physics of Astrophysics, Vol. II} (University Science
  Books)

\bibitem[{{Shu} et~al.(1987){Shu}, {Adams}, \& {Lizano}}]{1987ARA&A..25...23S}
{Shu}, F.~H., {Adams}, F.~C., \& {Lizano}, S. 1987, \araa, 25, 23

\bibitem[{{Solomon} et~al.(1987){Solomon}, {Rivolo}, {Barrett}, \&
  {Yahil}}]{1987ApJ...319..730S}
{Solomon}, P.~M., {Rivolo}, A.~R., {Barrett}, J., \& {Yahil}, A. 1987, \apj,
  319, 730

\bibitem[{{Stevens} et~al.(1992){Stevens}, {Blondin}, \&
  {Pollock}}]{1992ApJ...386..265S}
{Stevens}, I.~R., {Blondin}, J.~M., \& {Pollock}, A.~M.~T. 1992, \apj, 386, 265

\bibitem[{{Troland} \& {Crutcher}(2008)}]{2008ApJ...680..457T}
{Troland}, T.~H., \& {Crutcher}, R.~M. 2008, \apj, 680, 457. \eprint{0802.2253}

\bibitem[{{van Dishoeck} \& {Black}(1988)}]{1988ApJ...334..771V}
{van Dishoeck}, E.~F., \& {Black}, J.~H. 1988, \apj, 334, 771

\bibitem[{{van Dishoeck} \& {Blake}(1998)}]{1998ARA&A..36..317V}
{van Dishoeck}, E.~F., \& {Blake}, G.~A. 1998, \araa, 36, 317

\bibitem[{{V\'azquez-Semadeni}(1994)}]{1994ApJ...423..681V}
{V\'azquez-Semadeni}, E. 1994, \apj, 423, 681

\bibitem[{{V\'azquez-Semadeni}(2009)}]{2009arXiv0902.0820V}
--- 2009, ArXiv e-prints. \eprint{0902.0820}

\bibitem[{{V{\'a}zquez-Semadeni} et~al.(2010){V{\'a}zquez-Semadeni},
  {Col{\'{\i}}n}, {G{\'o}mez}, {Ballesteros-Paredes}, \&
  {Watson}}]{2010ApJ...715.1302V}
{V{\'a}zquez-Semadeni}, E., {Col{\'{\i}}n}, P., {G{\'o}mez}, G.~C.,
  {Ballesteros-Paredes}, J., \& {Watson}, A.~W. 2010, \apj, 715, 1302.
  \eprint{1001.0802}

\bibitem[{{V{\'a}zquez-Semadeni} et~al.(2009){V{\'a}zquez-Semadeni},
  {G{\'o}mez}, {Jappsen}, {Ballesteros-Paredes}, \&
  {Klessen}}]{2009ApJ...707.1023V}
{V{\'a}zquez-Semadeni}, E., {G{\'o}mez}, G.~C., {Jappsen}, A.,
  {Ballesteros-Paredes}, J., \& {Klessen}, R.~S. 2009, \apj, 707, 1023.
  \eprint{0904.4515}

\bibitem[{{V{\'a}zquez-Semadeni} et~al.(2007){V{\'a}zquez-Semadeni},
  {G{\'o}mez}, {Jappsen}, {Ballesteros-Paredes}, {Gonz{\'a}lez}, \&
  {Klessen}}]{2007ApJ...657..870V}
{V{\'a}zquez-Semadeni}, E., {G{\'o}mez}, G.~C., {Jappsen}, A.~K.,
  {Ballesteros-Paredes}, J., {Gonz{\'a}lez}, R.~F., \& {Klessen}, R.~S. 2007,
  \apj, 657, 870. \eprint{arXiv:astro-ph/0608375}

\bibitem[{{V{\'a}zquez-Semadeni} et~al.(2005){V{\'a}zquez-Semadeni}, {Kim},
  {Shadmehri}, \& {Ballesteros-Paredes}}]{2005ApJ...618..344V}
{V{\'a}zquez-Semadeni}, E., {Kim}, J., {Shadmehri}, M., \&
  {Ballesteros-Paredes}, J. 2005, \apj, 618, 344.
  \eprint{arXiv:astro-ph/0409247}

\bibitem[{{V\'azquez-Semadeni} et~al.(2000){V\'azquez-Semadeni}, {Ostriker},
  {Passot}, {Gammie}, \& {Stone}}]{2000prpl.conf....3V}
{V\'azquez-Semadeni}, E., {Ostriker}, E.~C., {Passot}, T., {Gammie}, C.~F., \&
  {Stone}, J.~M. 2000, Protostars and Planets IV, 3.
  \eprint{arXiv:astro-ph/9903066}

\bibitem[{{V\'azquez-Semadeni} et~al.(1995){V\'azquez-Semadeni}, {Passot}, \&
  {Pouquet}}]{1995ApJ...441..702V}
{V\'azquez-Semadeni}, E., {Passot}, T., \& {Pouquet}, A. 1995, \apj, 441, 702

\bibitem[{{V\'azquez-Semadeni} et~al.(1996){V\'azquez-Semadeni}, {Passot}, \&
  {Pouquet}}]{1996ApJ...473..881V}
--- 1996, \apj, 473, 881. \eprint{arXiv:astro-ph/9607046}

\bibitem[{{V{\'a}zquez-Semadeni} et~al.(2006){V{\'a}zquez-Semadeni}, {Ryu},
  {Passot}, {Gonz{\'a}lez}, \& {Gazol}}]{2006ApJ...643..245V}
{V{\'a}zquez-Semadeni}, E., {Ryu}, D., {Passot}, T., {Gonz{\'a}lez}, R.~F., \&
  {Gazol}, A. 2006, \apj, 643, 245. \eprint{arXiv:astro-ph/0509127}

\bibitem[{{Vishniac}(1994)}]{1994ApJ...428..186V}
{Vishniac}, E.~T. 1994, \apj, 428, 186. \eprint{arXiv:astro-ph/9306025}

\bibitem[{{Wang} et~al.(2010){Wang}, {Li}, {Abel}, \&
  {Nakamura}}]{2010ApJ...709...27W}
{Wang}, P., {Li}, Z., {Abel}, T., \& {Nakamura}, F. 2010, \apj, 709, 27.
  \eprint{0908.4129}

\bibitem[{{Wolfire} et~al.(1995){Wolfire}, {Hollenbach}, {McKee}, {Tielens}, \&
  {Bakes}}]{1995ApJ...443..152W}
{Wolfire}, M.~G., {Hollenbach}, D., {McKee}, C.~F., {Tielens}, A.~G.~G.~M., \&
  {Bakes}, E.~L.~O. 1995, \apj, 443, 152

\bibitem[{{Zuckerman} \& {Palmer}(1974)}]{1974ARA&A..12..279Z}
{Zuckerman}, B., \& {Palmer}, P. 1974, \araa, 12, 279

\end{thebibliography}

\end{document}